# Educational Virtual Field Trips based on Social VR and 360° Spaces

Surya Kalvakolu[1][0009-0001-5919-7253], Heinrich Söbke[1,2][0000-0002-0105-3126],
Jannicke Baalsrud Hauge[3,4][0000-0002-3747-0845], and Eckhard Kraft[1]

[1] Bauhaus-Universität Weimar, Goetheplatz 7/8, 99423 Weimar, Germany
{suryaprakash.reddy.kalvakolu@uni-weimar.de|heinrich.soebke|eckhard.kraft}@uni-weimar.de
[2] Hochschule Weserbergland, Am Stockhof 2, 31785 Hameln, Germany
soebke@hsw-hameln.de
[3] BIBA GmbH, Hochschulring 20, 28359 Bremen, Germany
baa@biba.uni-bremen.de
[4] Royal Institute of Technology, Kvarnbergagatan 12, Södertälje, Sweden
jmbh@kth.se

**Abstract.** Virtual field trips (VFTs) have proven to be valuable learning tools. Such applications are mostly based on 360° technology and are to be characterized as single-user applications in technological terms. In contrast, Social VR applications are characterized by multi-user capability and user-specific avatars. From a learning perspective, the concepts of collaborative learning and embodiment have long been proposed as conducive to learning. Both concepts might be supported using Social VR. However, little is currently known about the use of Social VR for VFTs. Accordingly, the research questions are to what extent VFTs can be implemented in Social VR environments and how these Social VR-based VFTs are perceived by learners. This article presents an evaluation study on the development and evaluation of a VFT environment using the Social VR platform Mozilla Hubs. It describes the design decisions to create the environment and evaluation results from a mixed-method study (N=16) using a questionnaire and focus group discussions. The study highlighted the opportunities offered by Social VR-based VFTs but also revealed several challenges that need to be addressed to embrace the potential of Social VR-based VFTs to be utilized regularly in education.

**Keywords:** social presence, collaborative learning, avatar, embodiment, virtual meetings, 360-degree

## 1 Introduction

Virtual field trips (VFTs) have proven to be powerful learning activities that promote exploratory learning through the visualization of real (physical) sites [1–3]. In the context of this article, the term VFT is defined as a VR space based on 360° panoramas that are visualized highly realistic [4, 5].
360° technology is a projection of spherical images to render the surroundings, where a camera view is positioned at the center of the image plane to realize human-eye behavior. The spherical images are generated either from machine-aided applications (e.g., building information modelling) or from photographic imagery. 360° panorama-based VFTs offer low-cost, easy-to-capture, non-computer-generated simulations that can provide a true-to-reality representation of the location [6]. The application of 360° imagery offers a unique sense of presence and immersion that has potential in teaching and learning [7].

VFTs do not fully achieve the authenticity and information content of field trips [8, 9], despite the evidence of the comparability of photos and reality [10]. On the other hand, there are advantages such as location and time independence, avoiding weather issues, overcoming physical inaccessibility, or avoiding dangerous places [11] as well as that they are less time and cost-consuming to use [1, 12, 13]. VFT environments are mostly single-user applications, which limits many learning processes, like situated learning that requires the collaborative solving of tasks [14]. Consequently, collaborative learning is also not inherently promoted [15]. One example in which collaborative work is achieved via organizational measures outside the virtual environment rather than via technological means is the implementation of the VFT as an escape room game [16]. Furthermore, a few prototypes are using multi-user environments for VFTs (e.g., [17]), but these are currently less openly accessible.

A type of virtual environment, which on the one hand uses the spatial metaphor, and on the other hand is multi-user capable by definition, is Social Virtual Reality, i.e. a multi-user virtual reality platform lets learners interact and communicate in virtual environments using avatars [18, 19]. Social VR allows learners to meet in virtual environments, participate in activities, and communicate. Further, it is considered essential for remote collaboration in virtual environments [20–23].

Combining these collaborative learning beneficial advantages of Social VR with the advantages of VFT seems therefore to be a promising approach.

Accordingly, this evaluation study is guided by the following research questions (RQ):

**RQ 1**: To what extent is it possible to implement a VFT environment with the help of the social VR platform Mozilla Hubs?

**RQ 2**: What are learners' attitudes towards such a VFT environment?

The rest of the article is structured as follows: In the next chapter, we elaborate on important aspects of the theoretical background. Chapter 3 explains the multi-stage study design, and Chapter 4 presents the results, which are then finally discussed in Chapter 5.

## 2      Theoretical Background

To our knowledge, Social VR has not been used to conduct VFTs. In the following, we argue why Social VR can support the implementation of VFTs. One of the important characteristics of Social VR is that it supports collaborative learning [24, 25]. Collaborative learning takes place in a situation in which two or more learners learn or attempt to learn something together to build a shared body of knowledge. In contrast to individual learning, learners participate in a collaborative learning process to benefit from each other's resources and competencies [26]. In other words, collaborative learning refers to methods and specific environments in which learners participate in a shared task in which learners are interdependent and responsible for each other [27].

Social VR platforms allow learners to interact with each other and possibly also with teachers in virtual environments using avatars. This promotes a sense of social presence and facilitates communication, interaction and collaboration, and therefore collaborative learning takes place [28–30]. Social presence is seen as being with real people in a virtual environment [31, 32]. Social presence enables multiple learners to communicate with each other, usually through visual and auditory cues [33, 34]. Social presence enables communicative behavior during virtual field trips. It is hypothesized that the stronger feeling of being physically present in the VFT the more it encourages learners to be more confident and intrinsically motivated to learn [35]. In addition, social presence contributes significantly to learner engagement in the virtual environment [36]. Accordingly, social VR-based learning experiences enable learners to observe other learners and subsequently emulate models for successful learning [37]. Since social presence is considered a prerequisite for learning [38], we consider the increase in social presence through Social VR to be an important argument in favor of Social VR-based VFTs.

A striking feature of social VR is the representation of participants through three-dimensional representations, so-called avatars. Participants can customize such avatars and control these avatars through the social VR environment. The use of avatars is one of the factors that increase social presence [39], especially in social VR environments [40]. Furthermore, the concept of Sense of Embodiment (SoE) to a body B is described by Kilteni et al. [41] concerning virtual reality as 'the sense that emerges when B's properties are processed as if they were the properties of one's own biological body'. Accordingly, we hypothesize possible effects of avatar embodiment, i.e. the conceptual extension of embodied cognition [42] beyond the learner's own body to its digital representation, the avatar. Although in this study both concepts, collaborative learning and avatar embodiment, are not investigated in detail, we believe that this study provides preliminary work for a more detailed investigation.

## 3      Methodology: Design of prototype and experimental set-up

The prototype was developed in line with the design guidelines in [43], starting with collecting requirements, then generating mock-ups (in our case scenarios), assessing these and then starting the validation process, which is described in more detail below.

The requirements for the prototype are based on a **brainstorming session** (results described in Chapter 3.1) among the authors. Based on these requirements, the first author developed a set of different scenarios including an assessment of implementation in various virtual environments leading to a **platform selection** (3.2). In the next step, the first author **developed the environment** (3.3) and documented his design choices as well as challenges. The prototype was finally validated in a study comprising 16 researchers and assistants (4 female, 12 male) recruited from a chair of waste management. The participants had an average age of 33.9 years (SD=9.65, range 21-56).

The **validation process** was carried out in the following way:

1. The participants were invited to an online meeting.

2. In the online meeting, the participants got an introduction to the virtual environment, further they watched an onboarding video on how to create their avatar and how to use the virtual environment.

2. The participant joined the virtual environment received an introduction to the navigation dashboard and moved to the area of the virtual lecture hall.

3. The facilitator took the role of guiding the participants through the lecture hall and the navigation dashboard.

4. Within a time frame of 15 minutes the participants explored the virtual environment and the participants were guided through the first four panoramas discussing the content of the panoramas (e.g., Fig. 3).

5. The participants completed the questionnaire, which also asked about their prior knowledge of the subject and media technology (Table 1). In addition, the demographic data already reported was requested. The questionnaire also consisted of three self-designed item sets with a total of 17 items on handling, participants' impressions and appropriateness for learning.

6. Finally, a focus group discussion was held for approximately 25 minutes involving lively contributions from the participants

**Table 1.** Self-reported prior knowledge
(10-point Likert scale 1: No knowledge at all – 10 International expert)

| Knowledge Area | Mean (SD) |
|---|---|
| Composting | 7.3 (2.17) |
| Avatars | 4.0 (2.65) |
| Virtual Reality | 5.3 (2.54) |

## 4 Results

This section describes the results both from the brainstorming session as well as the overall prototype-validation process.

### 4.1 Requirements

Within the **brainstorming session**, requirements for the virtual environment were identified. In this session, the participants achieved a mutual agreement on the suitability and advantage of including various features of Social VR for VFTs. In the following, the requirements are described.

- **Multi-user functionality** enables multiple learners to collaborate simultaneously within a virtual environment regardless of the geographical location.
- **Avatars and locations** provide learners with the flexibility to navigate and interact within the virtual environment. Learners can explore different locations in the virtual environment and interact within the virtual environment using their avatars.
- **Virtual meetings** support the organization of lectures, gatherings and discussions within the virtual environment. Communication is supported through various channels such as chat, voice, or video, facilitating effective collaboration and participation.
- **Screensharing and live cameras** facilitate the exchange of visual and audio content among learners. This feature enables presentations, demonstrations, and discussions, enhancing the collaborative experience within the virtual environment.
- **360°-Panoramas** are required for VFTs; thus, this needs to be supported by the selected platform.

### 4.2 Platform Selection

Different Social VR and similar platform (Artsteps, chilloutVR, Facebook Horizon, Mozilla Hubs, Mibo, Oasis, OpenSim, Overte, Second Life, Thinglink, VirBELA, vircadia and VRChat) were benchmarked against the above-mentioned criteria as well as the additional criteria of costs and developer community connections. This resulted in the selection of Mozilla Hubs for which a serviceable instance was already hosted by a partner institution (blinded for peer review). Accordingly, the prototype has been developed on Mozilla Hubs by using 360° panoramas.

### 4.3 Prototype Development

In professional 360° VR authoring systems, such as 3D Vista Virtual Tour Pro, the panoramas are stitched together to create a single walk-through space. Such a space cannot be created with Social VR software and Mozilla Hubs in particular. Accordingly, an alternative concept had to be found. The concept envisages that each panorama is placed as its own independent, self-contained world on a plane (Fig. 1). Learners can enter each panorama by teleporting themselves via a navigation board (Fig. 2, left), on which all panoramas are listed. The panorama can be exited again using the same mechanism in the direction of the navigation board by pressing a button; alternatively, the learner can continue directly to the previous or next panorama by pressing two further buttons - all panoramas may be visited in a defined sequence according to the metaphor of a field trip. The setting is complemented by a lecture hall in which learners can virtually sit down in front of a virtual monitor for the presentation of videos, slides or lectures (Fig. 1, top left).

During the VFT, learners can participate in public chat, fostering communication with fellow learners. Challenges arise when learners are distant in the virtual environment, affecting the clarity and audibility of their voices. To address this, a Megaphone feature has been incorporated by Mozilla Hubs, allowing learners to make announcements that are audible to learners at a distance. Additional features such as volume adjustment, screen, and camera sharing, incorporating of 3D models, pen usage, document sharing, and avatar customization are available. Finally, the prototype addresses the requirements and thus supports VFTs based on the following functionalities: interaction with the virtual environment, multi-user capability, and communication through different mediums such as microphones, text messages and documents [44, 45].

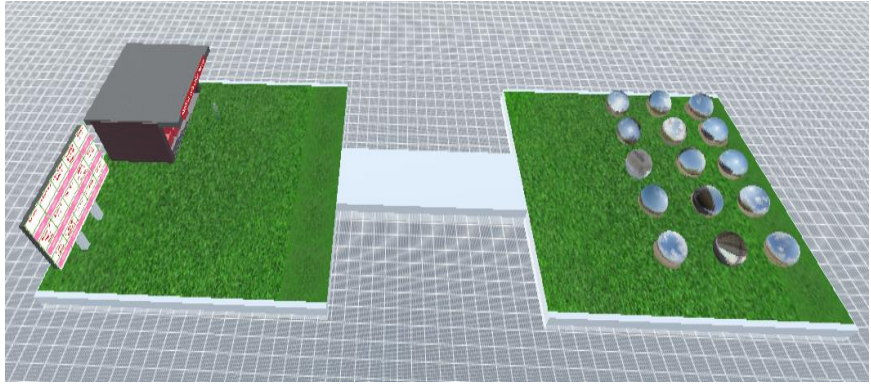

**Fig. 1.** Virtual environment: Lecture Hall and Navigation Board (left) and Panoramas (right)

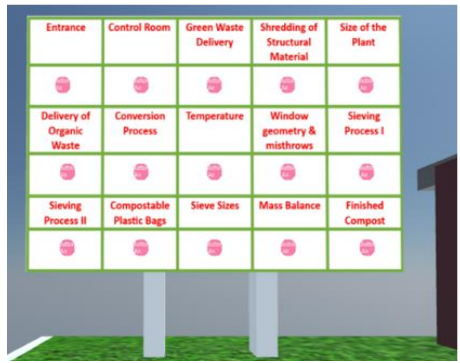

**Fig. 2.** Navigation Board

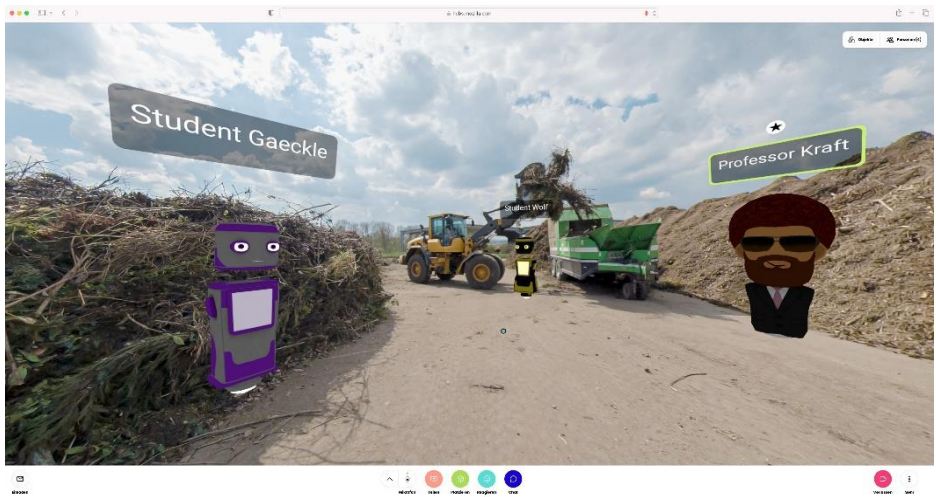

**Fig. 3.** Screenshot: Panorama and VFT participants

### 4.4 Development Challenges

While developing the virtual environment, the 3D models, such as the navigation board and the lecture hall have been developed in Blender and then has been imported into the Mozilla Hubs. Further challenges are clustered and described in the following.

**Spatial Environment. (I)** Avatars were able to move accidentally out of panoramas. Instead of using default spheres for displaying the panoramas, a mesh has been developed in Blender and used for providing a rigid body in Mozilla Hubs. **(II)** Other spheres turned out to be visible when learners take certain positions when they are in a sphere. This issue could be eliminated by placing walls around each sphere. **(III)** Avatars move in a circular motion. Accordingly, an external surface is introduced at a height from the surface to have a plane surface instead of a curved surface. By adopting the external surface, the plane movement of the avatar could be achieved (Fig 4).

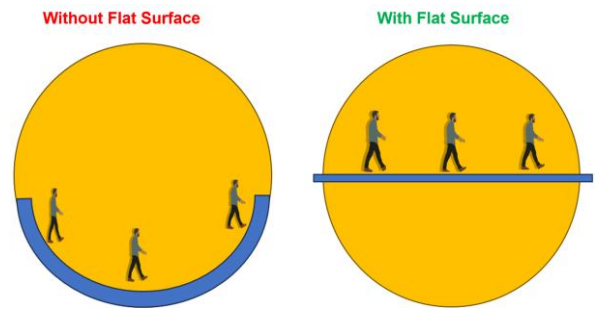

**Fig. 4.** Original (left) and flat surface

**User Support.** During a few pilot evaluations of the prototype, it was recognized that the handling of the virtual environment might overwhelm some learners. Accordingly, a video was recorded in which the handling of the virtual environment is explained. The video was used in the study.

### 4.5 User Experience Questionnaire

In line with the prototype character of the virtual environment and given the assumed low level of prior knowledge of VR and avatars, the items in the first item set on handling included questions on the ease of navigation (Fig. 5). More than 5 out of 7 points were scored for the actions of logging in, changing the panorama, going to a specific location and changing direction. That the participants were more concerned with themselves, and the handling of the virtual environment is indicated by the lower score (4.7) for paying attention to the guide. The virtual environment was also perceived as not too intuitive (4.5). The least effort (3.1) was put into customizing the avatar.

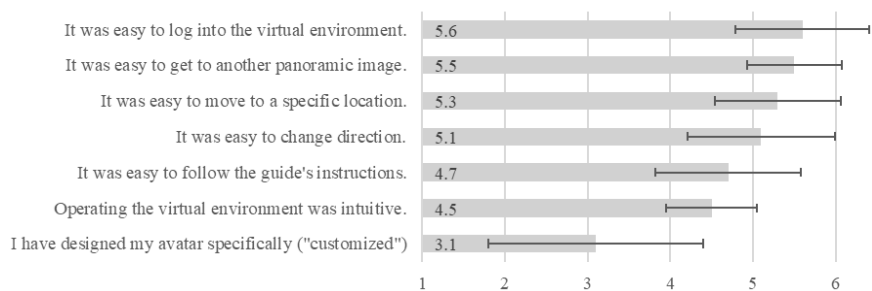

**Fig. 5.** Operation (7-point Likert scale, 1: I do not agree– 7: I agree, N=16)

The second item set (Fig. 6) asked about the participants' perceptions. The participants described the visual information of the panoramas as rather easy to grasp (5.1) although this was perceived less as natural (4.1). The acoustic clarity of the other participants received the second-highest score (4.7). There was no consensus on the question of the extent to which the handling of the virtual environment diverted attention away from the subject content (4.2). This value may also be explained in light that the virtual environment, but not the subject context, was more likely to be something new for the participants. The participants were also asked about the influence of other participants on their movements. This resulted in a score one point above the average (4.6). Given the panoramas described as narrow, this value appears to be moderate.

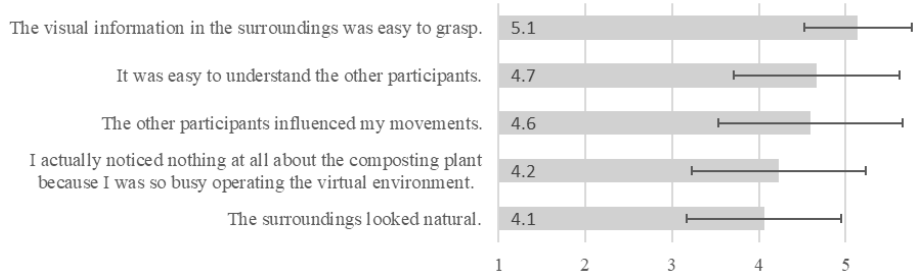

**Fig. 6.** Perceptions (7-point Likert scale, 1: I do not agree– 7: I agree, N=16)

Finally, questions were asked about the learning appropriateness of the environment (Fig. 7). Almost all the participants are to be considered specialists in teaching - either as lecturers or students - and their statements are correspondingly significant. The high value (5.6) for the desire to stay in the virtual environment is seen as a generally high level of interest. The teaching potential of the virtual environment is also recognized (5.4) and it is said to provide a realistic impression of the composting plant represented (5.3). The participants can also imagine using the application in their own teaching and learning scenarios. The item inserted as a test question, which sees the virtual environment as a technical gimmick with no great benefit for teaching, accordingly, receives the lowest and below-average approval (3.3)

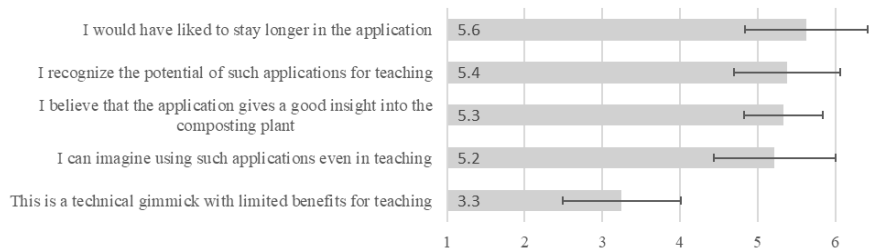

**Fig. 7.** Learning (7-point Likert scale, 1: I do not agree– 7: I agree), N=16

The questionnaire used an open question asking the participants about their experiences and needs for the virtual environment. A qualitative content analysis [47] of the 14 responses by two of the authors led to a total of 17 codes, which could be summarized in three clusters (Handling, Impressions, Features). The *Handling* cluster included the difficulty of handling the virtual world with the mouse, for example, the direction of movement was described as counterintuitive (2 mentions). Navigation was also described as not natural (1) and tedious (1). Onboarding, i.e. the introduction to the virtual environment, was criticized as having space for improvement. Furthermore, the navigation board was named as the preferred destination for teleporting from the panoramas; the current destination is the lecture hall. In the *Impressions* cluster, the most frequent comment was positive feedback about the virtual environment (3). It was criticized that too many people were competing for space in the panoramas at the same time (2). Accordingly, it was also noted that the focus was less on the subject content environment than on the virtual environment and the other avatars (1). Suggestions for improving the virtual environment were collected in the Features cluster. For example, it was requested that the avatars should be better labelled with (real) names (2). During the study, the name was placed rather far above the avatar, so that other participants who were close to the avatar had to change their line of vision. In addition, there were no instructions during onboarding for participants to customize their avatar with their real name. The possibility of further interaction between the avatars, which was not available in the virtual environment of the study, was also desired. Reference was made to the lack of annotations (1), i.e. additional information such as text, graphics and videos, which are essential in the 360°VR single-user applications. Furthermore, a better overview (1) was requested, i.e. the current position of the panorama should be made clear by employing a head-up mini-map, for example.

### 4.6   Focus Group Discussion

The focus group discussion, which was held after the questionnaires had been completed and had the same 16 participants, focused on the handling of, perceptions of and appropriateness of the virtual environment as a learning tool. In particular, the difficulties perceived by the participants are described below.
**Avatar Movement and Positioning.** Learners have reported difficulties inconsistently experiencing a sense of walking on the ground within the virtual environment. Learners find themselves either floating in the air or merging into the model ground. This persistent issue hampers the overall realism in the virtual environment.

**Avatar Identification and Interaction.** Learners face challenges in identifying and interacting with fellow learners, especially when they are in proximity or at a distance. The size and distance of the avatar heads associated nametag contribute to visibility issues.

**Space Constraints in Panoramas.** The size of panoramas in the virtual environment poses challenges related to the claustrophobic effect experienced when a considerable number of learners populate the shared space. Increasing the size of the sphere is a potential solution, but this is hindered by concerns related to image quality, necessitating a delicate balance between user comfort and visual fidelity.

**Shared Document Movement.** The collaborative feature allowing learners to view documents, such as notes, PDFs, and spreadsheets, presents challenges when multiple learners attempt to control the position of the document simultaneously. This results in a disorienting experience for learners, highlighting the need for improved mechanisms to manage shared documents within the virtual environment.

**Navigation Approach.** Some learners express a disconnect with the VFT within the virtual environment, emphasizing the absence of a natural walking sensation between locations on the virtual plane. The reliance on buttons for location transitions diminishes the sense of exploration, prompting a re-evaluation of the navigation approach to enhance the immersive nature of the VFT.

## 5    Discussion and Conclusions

Virtual Field Trips (VFTs) have proven to be powerful learning tools. In part, however, VFTs suffer from being single-user applications. Accordingly, they can only support learning-fostering approaches, such as situated learning or collaborative learning, to a limited extent. In this study, we have therefore attempted to combine 360°-based VFTs with the concept of Social VR: a VFT environment was created using the Social VR software Mozilla Hubs (RQ 1). The established spatial metaphor of Social VR needed to be adapted for the VFT environment: Each 360° panorama became its room within a room, in which learners can be teleported via buttons. In each 360° panorama itself, it is possible to teleport to the dashboard and to upstream and downstream panoramas in the context of the VFT. In general, we received a positive assessment of the potential of the proposed virtual environment in learning contexts (RQ 2). However, several challenges had to be overcome during the development of the virtual environment like the limited support for 360° panoramas, which resulted in a spherical, round floor area. Furthermore, before productive usage, it is necessary to look into aspects like the limitation in the number of learners in the panoramas due to little space. In addition, nametags are not placed so that they are recognizable, which means that social presence is not optimally promoted. In addition to name tags, it should also be investigated to what extent lifelike avatars lead to an increase in social presence. All in all, the experiences call for specialized authoring software for VFTs that supports inherently the concept of Social VR.

In addition to the small sample size of the evaluation, the limitations of this study include its status as an evaluation study that was able to test the simple feasibility of the conceptual approach. For a targeted use of these VFT environments in teaching, there is a lack of evidence-based knowledge about the extent to which the involvement of multiple learners and the representation by self-controlled avatars (avatar embodiment) are conducive to learning outcomes. Future work includes studies to validate the concepts of learning psychology, in particular social presence and avatar embodiment. Furthermore, interaction possibilities, especially between learners as well as between learners and the environment seem to be relevant to investigate. Overall, the study has confirmed the potential of a combination of VFTs and Social VR to promote learning and thus opens the way for a very promising and powerful learning tool.

## 6    References

1. Norris E, Shelton N, Dunsmuir S, et al (2015) Teacher and pupil perspectives on the use of Virtual Field Trips as physically active lessons. BMC Res Notes 8:719. https://doi.org/10.1186/s13104-015-1698-3
2. Seifan M, Dada D, Berenjian A (2019) The effect of a virtual field trip as an introductory tool for an engineering real field trip. Educ Chem Eng 27:6–11. https://doi.org/10.1016/j.ece.2018.11.005
3. Wolf M, Wehking F, Söbke H, et al (2023) Virtualised virtual field trips in environmental engineering higher education. Eur J Eng Educ 48:1312–1334. https://doi.org/10.1080/03043797.2023.2291693
4. Pham HC, Dao N, Pedro A, et al (2018) Virtual field trip for mobile construction safety education using 360-degree panoramic virtual reality. Int J Eng Educ 34:1174–1191
5. Ritter KA, Chambers TL (2022) Three-dimensional modeled environments versus 360 degree panoramas for mobile virtual reality training. Virtual Real 26:571–581. https://doi.org/10.1007/s10055-021-00502-9
6. Eiris R, Gheisari M, Esmaeili B (2020) Desktop-based safety training using 360-degree panorama and static virtual reality techniques: A comparative experimental study. Autom Constr 109:102969. https://doi.org/10.1016/j.autcon.2019.102969
7. Reyna J (2018) THE POTENTIAL OF 360-DEGREE VIDEOS FOR TEACHING, LEARNING AND RESEARCH. Valencia, Spain, pp 1448–1454

8. Springer C, Wehking F, Wolf M, Söbke H (2020) Virtualization of Virtual Field Trips. In: Proceedings of DELbA 2020. CEUR-WS, Online
9. Wolf M, Wehking F, Montag M, Söbke H (2021) 360°-Based Virtual Field Trips to Waterworks in Higher Education. Computers 10:118. https://doi.org/10.3390/computers10090118
10. Rossetti T, Hurtubia R (2020) An assessment of the ecological validity of immersive videos in stated preference surveys. J Choice Model 34:100198. https://doi.org/10.1016/j.jocm.2019.100198
11. Klippel A, Zhao J, Jackson KL, et al (2019) Transforming Earth Science Education Through Immersive Experiences: Delivering on a Long Held Promise. J Educ Comput Res 57:1745–1771. https://doi.org/10.1177/0735633119854025
12. McCaffrey KJW, Hodgetts D, Howell J, et al (2010) Virtual fieldtrips for petroleum geoscientists. Geol Soc Lond Pet Geol Conf Ser 7:19–26. https://doi.org/10.1144/0070019
13. Schulze DG, Rahmani SR, Minai JO, et al (2021) Virtualizing soil science field trips. Nat Sci Educ 50:e20046. https://doi.org/10.1002/nse2.20046
14. Lave J, Wenger E (1991) Situated learning: legitimate peripheral participation. Cambridge University Press
15. Dillenbourg P (1999) Collaborative learning: Cognitive and computational approaches. advances in learning and instruction series. ERIC
16. Wolf M, Montag M, Söbke H, et al (2024) Low-Threshold Digital Educational Escape Rooms Based on 360VR and Web-Based Forms. Electron J E-Learn 22:01–18
17. Eiris R, Gheisari M (2022) iVisit-Collaborate: Online Multiuser Virtual Site Visits Using 360-Degree Panoramas and Virtual Humans. In: Linner T, García de Soto B, Hu R, et al (eds) Proceedings of the 39th International Symposium on Automation and Robotics in Construction. International Association for Automation and Robotics in Construction (IAARC), Bogot�, Colombia, pp 276–282
18. Li J, Vinayagamoorthy V, Schwartz R, et al (2020) Social VR: A New Medium for Remote Communication and Collaboration. In: Extended Abstracts of the 2020 CHI Conference on Human Factors in Computing Systems. ACM, Honolulu HI USA, pp 1–8
19. Wang M (2020) Social VR : A New Form of Social Communication in the Future or a Beautiful Illusion? J Phys Conf Ser 1518:012032. https://doi.org/10.1088/1742-6596/1518/1/012032
20. Ahlers T, Bumann C, Kölle R, Lazović M (2022) Foreign Language Tandem Learning in Social VR: Conception, Implementation and Evaluation of the Game-Based Application *Hololingo!* -Com 21:203–215. https://doi.org/10.1515/icom-2021-0039
21. Barreda-Ángeles M, Horneber S, Hartmann T (2023) Easily applicable social virtual reality and social presence in online higher education during the covid-19 pandemic: A qualitative study. Comput Educ X Real 2:100024. https://doi.org/10.1016/j.cexr.2023.100024
22. De Simone F (2018) Measuring User Quality of Experience in Social VR systems. In: Proceedings of the 3rd International Workshop on Multimedia Alternate Realities. ACM, Seoul Republic of Korea, pp 25–26
23. Gunkel SNB, Prins M, Stokking H, Niamut O (2017) Social VR Platform: Building 360-degree Shared VR Spaces. In: Adjunct Publication of the 2017 ACM International Conference on Interactive Experiences for TV and Online Video. ACM, Hilversum The Netherlands, pp 83–84
24. Mystakidis S, Berki E, Valtanen J-P (2021) Deep and Meaningful E-Learning with Social Virtual Reality Environments in Higher Education: A Systematic Literature Review. Appl Sci 11:2412. https://doi.org/10.3390/app11052412
25. Rante H, Zainuddin MA, Miranto C, et al (2023) Development of Social Virtual Reality (SVR) as Collaborative Learning Media to Support Merdeka Belajar. Int J Inf Educ Technol 13:1014–1020. https://doi.org/10.18178/ijiet.2023.13.7.1900
26. Collis B, Moonen J (2005) Collaborative Learning in a Contribution-Oriented Pedagogy: In: Howard C, Boettcher JV, Justice L, et al (eds) Encyclopedia of Distance Learning. IGI Global, pp 277–283
27. González-Ruiz V, Ortega G, Garzón G, et al (2019) COLLABORATIVE PROJECT-BASED LEARNING: AN EXPERIENCE. IATED, pp 8631–8635
28. Dunmoye ID, Rukangu A, May D, Das RP (2024) An exploratory study of social presence and cognitive engagement association in a collaborative virtual reality learning environment. Comput Educ X Real 4:100054. https://doi.org/10.1016/j.cexr.2024.100054
29. Montagud M, Li J, Cernigliario G, et al (2021) Towards SocialVR: Evaluating a Novel Technology for Watching Videos Together
30. Van Brakel V, Barreda-Ángeles M, Hartmann T (2023) Feelings of presence and perceived social support in social virtual reality platforms. Comput Hum Behav 139:107523. https://doi.org/10.1016/j.chb.2022.107523
31. Biocca F, Harms C, Burgoon JK (2003) Toward a More Robust Theory and Measure of Social Presence: Review and Suggested Criteria. Presence Teleoperators Virtual Environ 12:456–480. https://doi.org/10.1162/105474603322761270
32. Oh CS, Bailenson JN, Welch GF (2018) A Systematic Review of Social Presence: Definition, Antecedents, and Implications. Front Robot AI 5:114. https://doi.org/10.3389/frobt.2018.00114
33. Churchill EF, Snowdon D (1998) Collaborative virtual environments: An introductory review of issues and systems. Virtual Real 3:3–15. https://doi.org/10.1007/BF01409793
34. Snowdon D, Churchill EF, Munro AJ (2001) Collaborative Virtual Environments: Digital Spaces and Places for CSCW: An Introduction. In: Churchill EF, Snowdon DN, Munro AJ (eds) Collaborative Virtual Environments. Springer London, London, pp 3–17
35. Cheng K-H, Tsai C-C (2019) A case study of immersive virtual field trips in an elementary classroom: Students' learning experience and teacher-student interaction behaviors. Comput Educ 140:103600. https://doi.org/10.1016/j.compedu.2019.103600
36. Lee KM, Jung Y, Kim J, Kim SR (2006) Are physically embodied social agents better than disembodied social agents?: The effects of physical embodiment, tactile interaction, and people's loneliness in human–robot interaction. Int J Hum-Comput Stud 64:962–973. https://doi.org/10.1016/j.ijhcs.2006.05.002
37. Morrison L, Robb J, Hughes J, Lam M (2021) Social Presence in Virtual Professional Learning. J Digit Life Learn 1:93–110. https://doi.org/10.51357/jdll.v1i1.160

38. Richardson JC, Maeda Y, Lv J, Caskurlu S (2017) Social presence in relation to students' satisfaction and learning in the online environment: A meta-analysis. Comput Hum Behav 71:402–417. https://doi.org/10.1016/j.chb.2017.02.001
39. Yoon B, Kim H, Lee GA, et al (2019) The effect of avatar appearance on social presence in an augmented reality remote collaboration. IEEE, pp 547–556
40. Heidicker P, Langbehn E, Steinicke F (2017) Influence of avatar appearance on presence in social VR. In: 2017 IEEE Symposium on 3D User Interfaces (3DUI). IEEE, Los Angeles, CA, USA, pp 233–234
41. Kilteni K, Groten R, Slater M (2012) The Sense of Embodiment in Virtual Reality. Presence Teleoperators Virtual Environ 21:373–387. https://doi.org/10.1162/PRES_a_00124
42. Foglia L, Wilson RA (2013) Embodied cognition. WIREs Cogn Sci 4:319–325. https://doi.org/10.1002/wcs.1226
43. Camburn B, Viswanathan V, Linsey J, et al (2017) Design prototyping methods: state of the art in strategies, techniques, and guidelines. Des Sci 3:e13. https://doi.org/10.1017/dsj.2017.10
44. Eriksson T (2021) Failure and Success in Using Mozilla Hubs for Online Teaching in a Movie Production Course. In: 2021 7th International Conference of the Immersive Learning Research Network (iLRN). IEEE, Eureka, CA, USA, pp 1–8
45. Guichet PL, Huang J, Zhan C, et al (2022) Incorporation of a Social Virtual Reality Platform into the Residency Recruitment Season. Acad Radiol 29:935–942. https://doi.org/10.1016/j.acra.2021.05.024
46. Large DR, Hallewell M, Briars L, et al Pro-Social Mobility: Using Mozilla Hubs as a Design Collaboration Tool
47. Mayring P, Fenzl T (2019) Qualitative inhaltsanalyse. Springer